\documentclass[showpacs,showkeys]{revtex4}
\usepackage{graphicx}
\usepackage{amsmath}

\newcommand{\be}{\begin{equation}}
\newcommand{\ee}{\end{equation}}
\newcommand{\bea}{\begin{eqnarray}}
\newcommand{\eea}{\end{eqnarray}}

\begin{document}
\title{ Two flavor massless Schwinger model on a torus at a finite chemical potential
}
\date{\today}

\author{R. Narayanan}
\email{rajamani.narayanan@fiu.edu}
\affiliation{Department of Physics, Florida International University, Miami,
FL 33199.}

\begin{abstract}
We study the thermodyanamics of the two flavor massless Schwinger
model on a torus at a finite chemical potential. We show that the
physics only depends on the iso-spin chemical potential and there are
marked deviations from a free fermion theory. We argue
that spatial inhomogeneties can develop in the system at very low
temperatures. 
\end{abstract}

\maketitle

\section{Introduction}

A study of QCD in the presence of a finite chemical potential is
important
for our understanding of quark matter at finite density~\cite{Rajagopal:2000wf}. Lattice QCD
provides a non-perturbative approach to this problem and there
has been extensive work performed on this
topic~\cite{deForcrand:2010ys,Lombardo:2008sc}.
The fermion determinant in a fixed gauge field background is complex
(it can be made real by summing over a pair of gauge field and its complex conjugate
but the result is not necessarily positive) and therefore suffers from
the so-called ``sign problem''. QCD wth a finite isospin chemical
potential does not suffer from the ``sign problem'' and the physics of
this model has been explored~\cite{Son:2000xc,Detmold:2012wc}. 

The Schwinger model (QED in two dimensions) has played the role of a
very useful toy model for QCD in four dimensions.
Generalized Thirring models have been studied in detail at finite
temperature and finite chemical
potential~\cite{Sachs:1993zx,Sachs:1995dm}. 
One main result at is applies to the Schwinger model is the
independence on the chemical potential. This can be seen
as a consequence of the integral over toron fields
\cite{Langfeld:2011rh,Narayanan:2012du}
in the path integral formalism. The issue at hand is imposing Gauss's
law in the path integral formalism~\footnote{I would like to thank
  Philippe de Forcrand for pointing this out to me.}. Imposing Gauss's
law
in the Hamiltonian formalism results in the condition that the
time-like
component of the electromagnetic potential vanish at spatial
infinity~\cite{Gross:1980br}.
This amounts to setting the toron field in one direction on the torus
to zero in the path integral~\cite{Bender:1992gn}. This would allow
for states with net total charge to be present but would break the
$U(1)$ global symmetry associated with the Polyakov loop in the
time-like direction placing the theory in a deconfined phase.
We will study the two flavor massless Schwinger model on a finite
torus
in the presence of a chemical potential. We will integrate over the
toron fields. As expected, the theory will
be independent of the chemical potential that couples to the total
charge
but will depend on the isospin chemical potential.

We start with a definition of the model on a finite torus and state
the result for the fermion determinant in a fixed gauge field
background
using the zeta-function regularization~\cite{Sachs:1991en}. 
We will address the
integration of the fermion determianant over the toron fields and show that the result is
independent of the chemical potential that couples to the total
charge. 
We will proceed to address the physics of the isospin chemical
potential.

\section{The grand canonical partition function}

\subsection{Model basics}

Let $l$ be the circumference of the spatial circle and let $\beta$ be
the inverse temperature. We will use $l$ to set all scales in the
theory and define $\tau=\frac{l}{\beta}$ as the dimensionless
temperature. The physical gauge coupling is set to $\frac{e}{l}$ where
$e$ is dimensionless.

The Hodge decomposition of the U(1) gauge field on a $l\times\beta$
torus is
\bea
A_1(x_1,x_2) &=& \frac{2\pi h_1}{l} +\partial_1\chi(x_1,x_2)
- \partial_2\phi(x_1,x_2) - \frac{2\pi k}{l\beta} x_2 \cr
A_2(x_1,x_2) &=& \frac{2\pi h_2}{\beta} +\partial_2\chi(x_1,x_2)
+ \partial_1\phi(x_1,x_2),
\eea
where $-\frac{1}{2}\le h_\mu < \frac{1}{2}$ are the toron fields and $\chi(x_1,x_2)$
generates gauge transformations.
The electric field density is
\be
E(x_1,x_2) = \frac{2\pi k}{l\beta} + \partial^2 \phi(x_1,x_2),
\ee
where $\phi(x_1,x_2)$ is a periodic function on the torus with no zero
momentum mode
and $k$ is the integer valued topological charge.
The gauge action is 
\be
S_g = \frac{2\pi^2 \tau k^2}{ e^2} + \frac{l^2}{2e^2} \int d^2x
(\partial^2 \phi)^2.
\ee

The determinant of a massless Dirac fermion is zero unless $k=0$ and the
determinant for $k=0$ using zeta function regularization is~\cite{Sachs:1991en,Sachs:1995dm}
\be
Z_f(\phi,h_\mu,\mu_i,q_i) = e^{\frac{q_i^2}{2\pi} \int d^2x
  \phi \partial^2 \phi } \frac{1}{\eta^4(\tau)}
\sum_{n_1,n_2=-\infty}^\infty 
e^{-\pi \tau \left(n_1 + q_ih_2 - i\frac{\mu_i}{\tau}\right)^2}
e^{-\pi \tau \left(n_2 + q_ih_2 - i\frac{\mu_i}{\tau}\right)^2}
e^{2\pi i q_ih_1\left(n_1 - n_2\right)},\label{znk0}
\ee
where 
$q_i$ is the integer valued charge of the fermion and
$\frac{2\pi \mu_i}{l}$ is the chemical potential.
The Dedekind eta function, $\eta(\tau)$ is given by
\be
\eta(\tau)= e^{-\frac{\pi\tau}{12}} \prod_{n=1}^\infty \left( 1-
  e^{-2\pi\tau n}\right).
\ee
There is an infinite normalization factor that has been removed by
zeta function regularization from
the above formula. This factor does depend on $\tau$ but only in
a trivial manner as to shift the zero point energy. 

We define the Fourier components of $\phi(x_1,x_2)$ according to
\be
\phi(x_1,x_2) = \frac{e  }{4\pi^2
  \tau^{\frac{3}{2}}}{\sum'}_{k_1,k_2=-\infty}^\infty e^{\frac{2\pi
    i}{\beta}\left(\frac{k_1}{\tau}x_1 +
    k_2x_2\right)}\tilde\phi(k_1,k_2),
\ee
with
$\tilde\phi(-k_1,-k_2)=\tilde\phi^*(k_1,k_2)$ and the prime over sum
implies that $k_1=k_2=0$ is excluded.
Then
\be
\frac{l^2}{2e^2} \int d^2x (\partial^2\phi)^2 = \frac{1}{2}
{\sum'}_{k_1,k_2=-\infty}^\infty
\left | \phi(k_1,k_2) \right |^2 \left(k_2^2 + \frac{1}{\tau^2}
  k_1^2\right)^2,
\ee 
and
\be
\frac{q_i^2}{2\pi } \int d^2x \phi (\partial^2\phi) = \frac{e^2q_i^2}{8\pi^3\tau^2}
{\sum'}_{k_1,k_2=-\infty}^\infty
\left | \phi(k_1,k_2) \right |^2 \left(k_2^2 + \frac{1}{\tau^2}
  k_1^2\right).
\ee

\subsection{Bosonic and toronic partition functions}

We will consider the two flavor Schwinger model with $q_1=q_2=1$. 
We write the  partition function as
\be
Z(\mu_1,\mu_2,\tau,e) = Z_b(\tau,e) Z_t(\mu_1,mu_2,\tau)
\ee
where 
the first factor is the bosonic ($\phi$) partion function and the second
factor is the toronic ($h_\mu$) partition function. 

Since the total action (gauge and fermionic contribution) is a quadratic function in
$\phi$, the bosonic partition function is
\be
Z_b(\tau,e) = {\prod'}_{k_1,k_2=-\infty}^\infty
\sqrt{\frac{1}{\left(k_2^2+\frac{1}{\tau^2}k_1^2\right)
\left( k_2^2+\frac{1}{\tau^2}\left[k_1^2 +
    \frac{e^2}{2\pi^3}\right]\right)}}
\label{pfb}
\ee

Starting from (\ref{znk0}) and after a little bit of algebra,
the toronic partition function 
can be reduced to
\be
Z_t(\mu_1,\mu_2,\tau)
=
\eta^{-4}(\tau) \sum_{m_1,m_2,m_3=-\infty}^\infty
\int_{-\frac{1}{2}}^{\frac{1}{2}} dh_2  
e^{-\pi\tau\left[ 
\left(m_2-m_1 +2m_3+2h_2 -i\frac{\mu_1+\mu_2}{\tau}\right)^2
+\left(m_1 -i\frac{\mu_1-\mu_2}{\tau}\right)^2 
+m_2^2 \right]}.
\ee
Consider the integral,
\be
Z_3\left(k,\frac{\mu_1+\mu_2}{\tau}\right) =
\sum_{m_3=-\infty}^\infty
\int_{-\frac{1}{2}}^{\frac{1}{2}} dh_2  
e^{-\pi\tau
\left(k +2m_3+2h_2 -i\frac{\mu_1+\mu_2}{\tau}\right)^2}
\ee
Viewing, $z=h_2-i\frac{\mu_1+\mu_2}{2\tau}$, as a complex variable, we
see that the integrand is analytic in $z$ and periodic under $z\to
z+1$. Therefore, the integral is independent of
$\frac{\mu_1+\mu_2}{\tau}$. We explicitly see that the partition
function does not depend on the chemical potential that couples to the
total charge.
Note that the integrand is positive
definite, if we set $(\mu_1+\mu_2)=0$ but is not the case for a
general
$(\mu_1+\mu_2)$. One will encounter a {\sl sign problem} if one
tries
to compute the integral numerically with $(\mu_1+\mu_2)$ not equal to zero.
The integral is the same for all even $k$ and the same
for all odd $k$. We can write the reduced integral as
\be 
Z_3^k(\tau) = \sum_{m_3=-\infty}^\infty \int_{-\frac{1}{2}}^{\frac{1}{2}} dh_2 e^{-\pi \tau
    \left(k +2m_3 + 2h_2\right)^2}; \ \ \ k=0,1,
\ee 
Setting the dimensionless isospin chemical potential equal to 
\be
\mu_I = 2\pi(\mu_1-\mu_2),\label{muidef}
\ee
we have
\be
Z_t(\mu_I,\tau)
=\eta^{-4}(\tau)
e^{\frac{\mu_I^2}{4\pi \tau}}
\sum_{m_1,m_2=-\infty}^\infty \cos\left( m_1 \mu_I\right)
e^{-\pi \tau \left(m_1^2+m_2^2\right)} 
Z_3^{{\rm mod}(m_2-m_1,2)}(\tau).
\ee
Let
\bea
Z_2^0(\tau) &=& 
\sum_{m_2=-\infty}^\infty 
\left[ e^{-\pi\tau  (2m_2)^2} Z_3^0(\tau)
+e^{-\pi\tau  (2m_2+1)^2} Z_3^1(\tau)\right];\cr 
Z_2^1(\tau) &=& \sum_{m_2=-\infty}^\infty 
\left[ e^{-\pi\tau  (2m_2+1)^2} Z_3^0(\tau)
+e^{-\pi\tau  (2m_2)^2} Z_3^1(\tau)\right].
\eea
The final expression for the toronic partition function is
\be
Z_t(\mu_I,\tau)=
\eta^{-4}(\tau) e^{\frac{\mu_I^2}{4\pi\tau}}
\sum_{k=0}^1
\sum_{m_1=-\infty}^\infty \cos\left( \left(2m_1+k\right)\mu_I\right)
 e^{-\pi\tau  (2m_1+k)^2} Z_2^k(\tau).
\ee

\subsection{Thermodynamic observables}

The only contribution to the isospin number comes from the toronic
partition function and is
\be
N_I = \tau\frac{\partial \ln Z(\mu_I,\tau,e)}{\partial
  \mu_I} = \frac{\mu_I}{2\pi} -f(\mu_I,\tau),\label{density}
\ee
where
\be
f(\mu_I,\tau)=\tau
\frac
{
\sum_{k=0}^1
\sum_{m_1=-\infty}^\infty \left(2m_1+k\right)\sin\left(\left(2m_1+k\right)\mu_I\right)
 e^{-\pi\tau  (2m_1+k)^2} Z_2^k(\tau)
}
{
\sum_{k=0}^1
\sum_{m_1=-\infty}^\infty \cos\left( \left(2m_1+k\right)\mu_I\right)
 e^{-\pi\tau  (2m_1+k)^2} Z_2^k(\tau)
}. \label{density1}
\ee

The dimensionless energy is 
\be
E(\tau;N_I,e) = -\frac{\partial \ln
  Z(\mu_I,\tau,e)}{\partial{\frac{1}{\tau}}}\Bigg|_{\frac{\mu_I}{\tau}}
= E_b(\tau;e) + E_t(\tau;N_I)
\ee
and the contributions from the bosonic partition function and the
toronic partition function are
written separately.
The result from the toronic partition function is
\be
\frac{E_t(\tau;N_I)-E_t(0;0)}{\pi} 
=\frac{1}{3}\left(\tau^2+1\right) -\sum_{n=1}^\infty \frac{8
  n\tau^2}{e^{2\pi n\tau}-1}
+N_I^2 -f^2(\mu_I,\tau) +g(\mu_I,\tau)\label{etoron}
    \ee
where
\be
g(\mu,\tau)=
\tau^2\frac
{
\sum_{k=0}^1
\sum_{m_1=-\infty}^\infty \cos\left(\left(2m_1+k\right)\mu_I\right)
 \frac{d}{d(\pi\tau)}\left[e^{-\pi\tau  (2m_1+k)^2} Z_2^k(\tau)\right]
}
{
\sum_{k=0}^1
\sum_{m_1=-\infty}^\infty \cos\left( \left(2m_1+k\right)\mu_I\right)
 e^{-\pi\tau  (2m_1+k)^2} Z_2^k(\tau)
}.
\ee
The result from the bosonic partition function is
\be
E_b(\tau;e) -E_b(0;e)
=
\tau -\frac{e}{\sqrt{2\pi}} \left[\tanh \frac{e}{\sqrt{2\pi}\tau}-1\right]
-4\pi\sum_{k_1=1}^\infty k_1 \left[\tanh \frac{\pi k_1}{\tau}-1\right].\label{eboson}
\ee

\subsection{Free fermions}

In order to understand the results for the two flavor massless
Schwinger model, it is
useful to recall that the partition function for free fermions in one
dimension is given by
\be
\ln Z_f= \frac{2l}{\pi} \int_0^\infty dp \left [ \beta p + 
\ln \left(1+e^{-\beta(p-\mu_f)}\right) +
\ln \left(1+e^{-\beta(p+\mu_f)}\right) \right],
\ee
where $\mu_f$ is the chemical potential for free fermions
which we set to $\frac{\mu_I}{2l}$ in order to be consistent with the
two flavor notation in (\ref{muidef}).
The free fermion isospin number is given by
\be
N_f=\tau \frac{\partial \ln Z_f}{\partial \mu_I}\Bigg|_{\beta}
= \frac{\mu_I}{2\pi}.\label{fdensity}
\ee
After subtracting the zero point energy, the dimensionless energy of free fermions
at low temperatures is given by
\be
\frac{E_f(\tau;N_f) -E_f(0,0)}{\pi}= N_f^2+ \frac{1}{3}\tau^2 + \cdots\label{efree}
\ee
The first term is the fermi energy that grows quadratically with the
isospin number
and second term is the leading order low temperature correction that
is positive and quadratic in the temperature.

\section{Results and discussion}

\begin{figure}
\centerline{\includegraphics[scale=0.75]{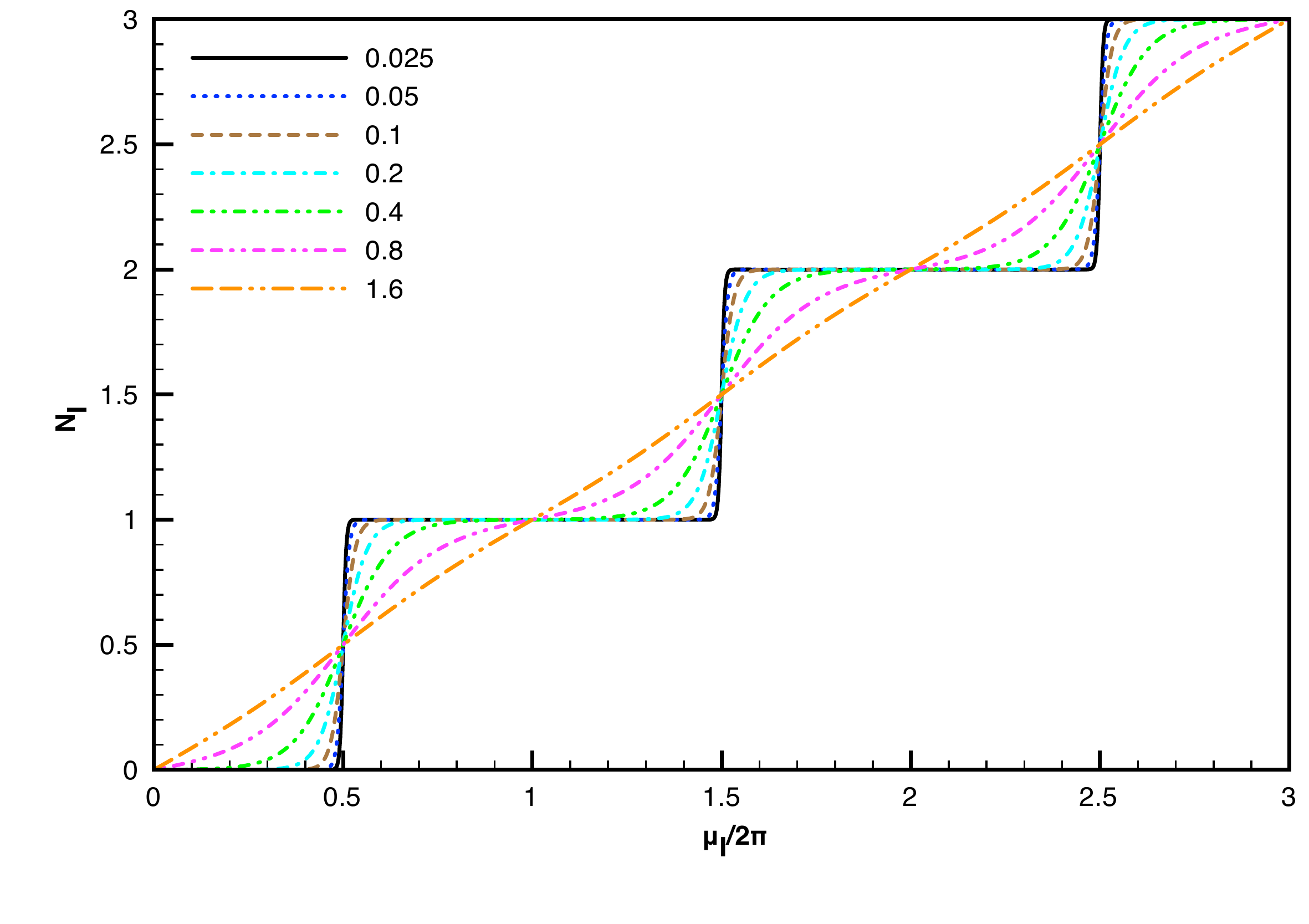}}
\caption{\label{fig1}
Plot of $N_I(\mu_I,\tau)$ versus $\mu_I$
(c.f.(\ref{density})) for several different values of $\tau$.
}
\end{figure}

We proceed to compare the results for the
two flavor Schwinger model with that for free fermions.
The result for the isospin number in (\ref{density}) is plotted
in Fig.~\ref{fig1} for several different values of temperature.
The linear behavior in (\ref{fdensity}) expected of two flavors of
free fermions in (\ref{fdensity}) 
is the first term in (\ref{density}) and this is achieved only in the
high temperature limit where the contribution from the second term
goes to zero. 
The first term is the na\"ive contribution from two flavors of free
fermions. 
The second term
is the result of integrating the effect of boundary conditions over
all
possible choices. 

\begin{figure}
\centerline{\includegraphics[scale=0.75]{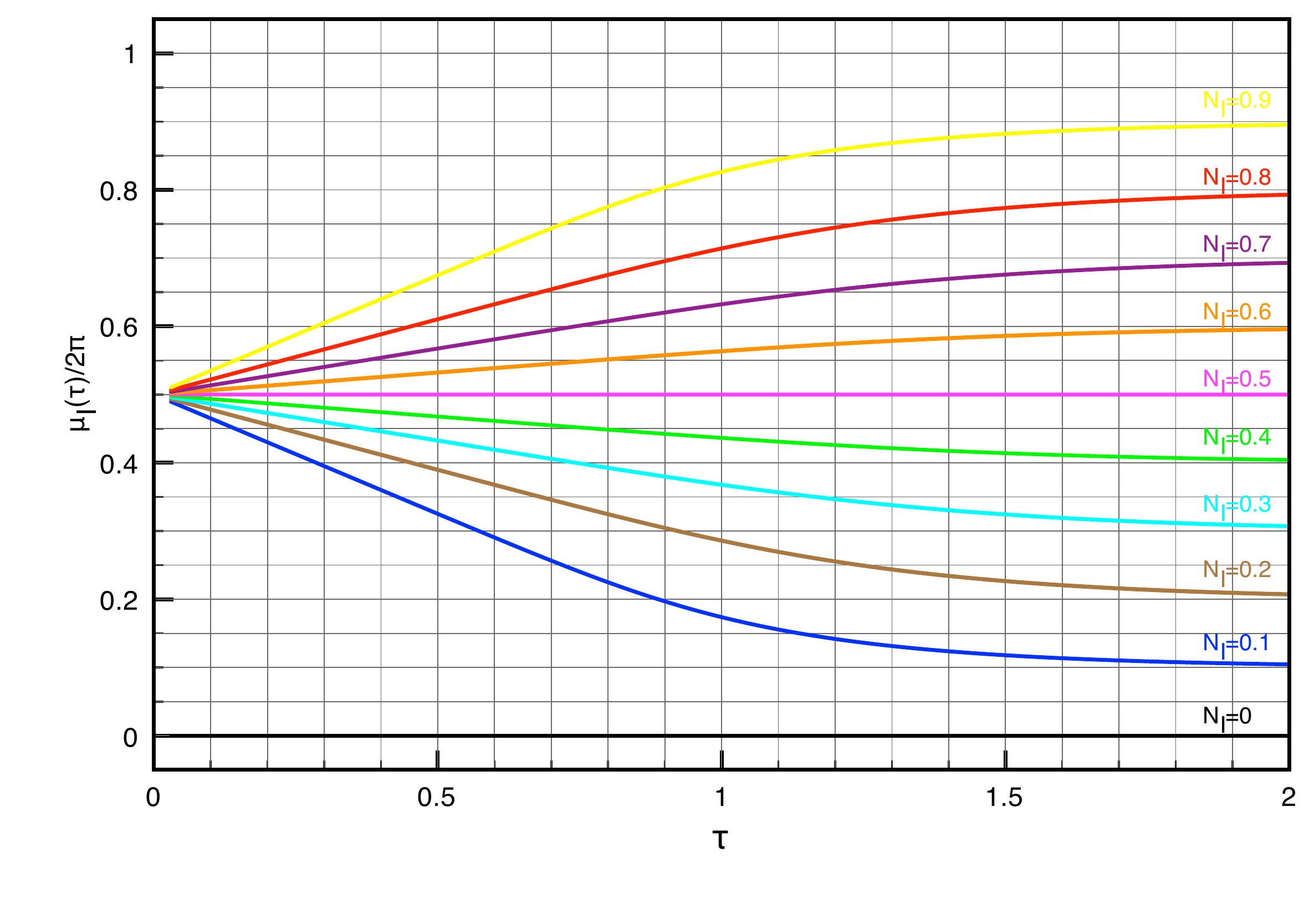}}
\caption{\label{fig2}
Plot of $\mu_I(\tau)$ versus $\tau$
(c.f.(\ref{density})) for several different values of $N_I$ in steps
of
$0.1$ starting from $0$ and ending in $3$.
}
\end{figure}

We can use Fig.~\ref{fig1} to see how the isospin chemical potential
depends on the temperature at a fixed isospin number. The quasi-periodicity
seen in the figure is a consequence of 
$f(\mu_I+2\pi,\tau)=f(\mu_I,\tau)$ 
in (\ref{density1}).
Furthermore, $f(\mu_I,\infty)=0$, and 
\be
\frac{\mu_I(\infty)}{2\pi}=N_I,\label{muinf}
\ee
  like for free fermions. On the other hand, 
\be
\lim_{\tau\to 0} f(\mu_I,\tau) = 
\begin{cases}
\frac{\mu_I}{2\pi} & \ \ {\rm if}\ \  0 < \mu_I < \pi \cr
\frac{\mu_I}{2\pi} -1 & \ \ {\rm if}\ \  \pi < \mu_I < 2\pi
\end{cases}.
\ee
Therefore,
\be
\frac{\mu_I(0)}{\pi} = \lceil{N_I}\rceil,\label{muzero}
\ee
for all non-integer values of $N_I$ and
$\lceil{N_I}\rceil$ is the ceiling function.
The behavior for non-zero and finite temperatures is to interpolate
between (\ref{muinf}) and (\ref{muzero}) as shown in Fig.~\ref{fig2}.
The behavior is shown for values of $N_I$ in the range $0\le N_I \le
3$
in steps of $0.1$. Plots are color coded to show periodicity of
$f(\mu_I,\tau)$.
Since $f(2n\pi,\tau)=0$ for all values of $\tau>0$ and any integer
$n$, we see that integer values of $N_I$ are special and behave like
free
fermions for all temperatures.

Since the partition function is independent of $(\mu_1+\mu_2)$, the
net charge is zero. But, we can maintain the system at a non-zero
isospin number, $N_I$. Since the system can exchange particles with
the reservoir, the expectation value of the isospin number need not be
an integer. Let us assume we start at high temperature with
a fixed $N_I$ assumed to take an integer value. Since the chemical
potential
does not change with temperature for this case and remains the unique
value for this particular value of $N_I$, the system will remain
homogeneous
at all temperatures. Now consider non-integer values of $N_I$.
Th system will be homogeneous at high temperatures since the chemical
potential is different for different values of $N_I$. As the system is
cooled and brought down to zero temperature, different values of $N_I$
can coexist as long as the different values all have the same ceiling
value, $\lceil N_I \rceil$ since they all have the same chemical
potential at zero temperature (c.f.(\ref{muzero})). The system is
bound to form inhomogeneities at zero temperature. 

We now proceed to use (\ref{density}) and (\ref{etoron}) to compute
the toronic contribution to the energy as a function of the
temperature at fixed isospin number. Consider the zero temperature
limit
in order to extract the fermi energy as a function of the iso-spin
number.
Since $f(2n\pi,0)$ and
$g(2n\pi,0)$ are zero it follows that the fermi energy for integer
values of isospin are given by the free fermion value.
As $\tau$ goes to zero,
$g(\mu,\tau)$ approaches a non-zero limit as long as $N_I$ is not an
integer. As a consequence, 
the fermi energy is given by
\be
E_F(N_I) = {\lfloor N_I\rfloor}^2 +
\left(2\lfloor N_I\rfloor+1\right)
\left(N_I - \lfloor N_I\rfloor\right),
\ee
and it linearly interpolates between the free fermion values at
integer values of $N_I$. 
We were unable to
  analytically obtain an explicit expression for the linear coefficient
  in (\ref{etoron}). We numerically evaluated it and found
  that the second term in (\ref{etoron}) contributes $\frac{2\tau}{\pi}$,
  and
the last two terms in (\ref{etoron}) contribute $-\frac{3\tau}{2\pi}$.
The leading behavior of the toronic contribution to the energy at low
temperature is
\be
\frac{E_t(\tau;N_I)-E_t(0;0)}{\pi} = 
E_F(N_I)
+ \frac{\tau}{2\pi}+\cdots.\label{lowtener}
\ee
This is qualitatively different from the free fermion result where the
leading term is quadratic in $\tau$. The linear coefficient of
$\frac{1}{2\pi}$
in (\ref{lowtener}) gets modified to $\frac{3}{2\pi}$ for the total
energy when we include the leading contribution from the bosonic 
partition function in (\ref{eboson}).

\begin{figure}
\centerline{\includegraphics[scale=0.75]{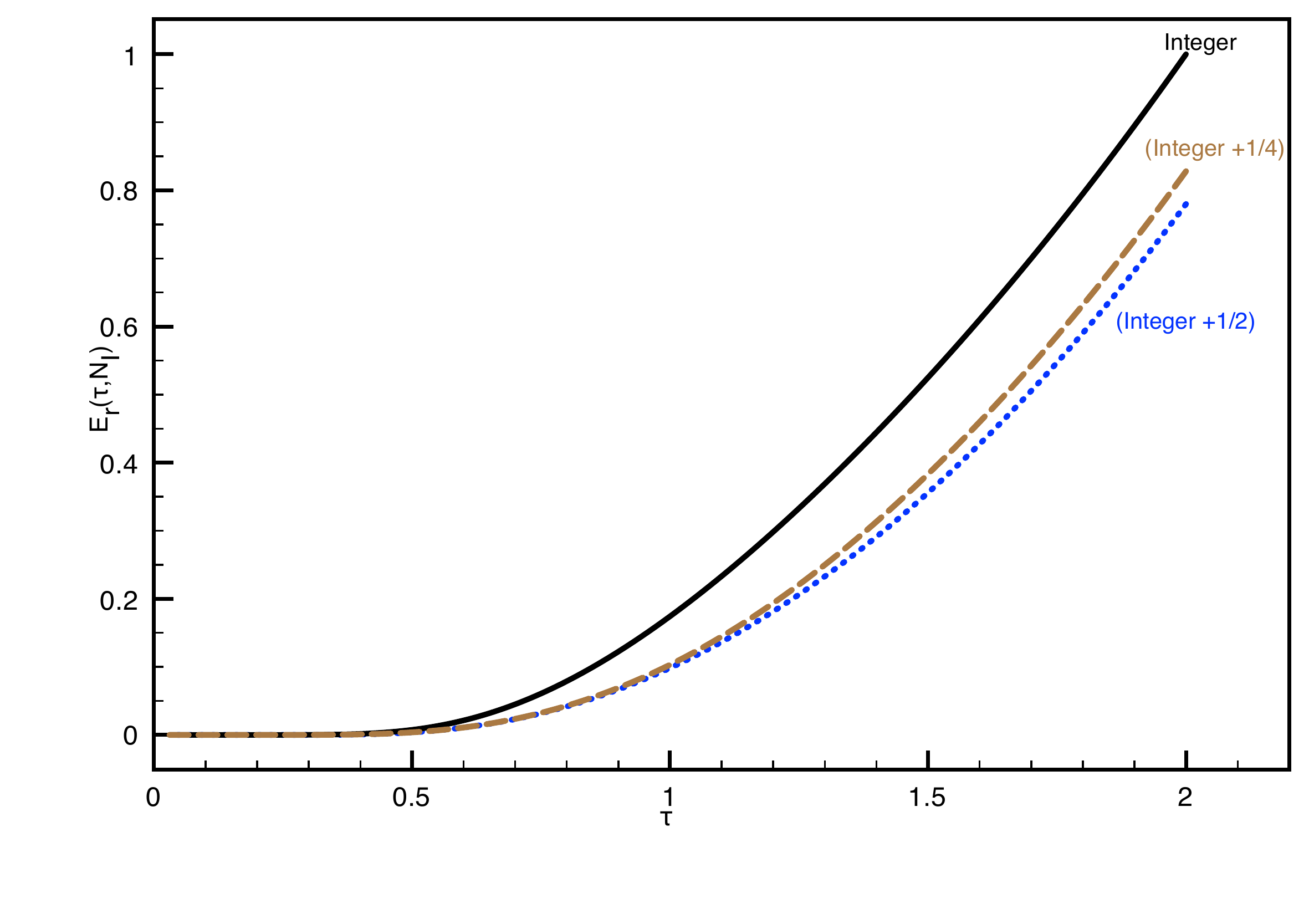}}
\caption{\label{fig3}
Plot of $E_r(\tau;N_I)$ versus $\tau$
(c.f.(\ref{erem})) for $N_I=0.1,0.2,0.3,0.4,0.5$. The color coding
is same as Fig.~\ref{fig2}.
}
\end{figure}

The higher order corrections in $\tau$ to the energy from the toronic
partition function,
\be
E_r(\tau,N_I) = \frac{E_t(\tau;N_I)-E_t(0;0)}{\pi} -
E_F(N_I)
- \frac{\tau}{2\pi},\label{erem}
\ee
is plotted as a function of $\tau$ for various values of $N_I$ in
Fig.~\ref{fig3} for $N_I=0.1,0.2,0.3,0.4,0.5$. Due to
quasi-periodicity,
$E_r(\tau,1-N_I)=E_r(\tau,N_I)$ for $(0 < N_I \le 0.5)$. 
In addition, $E_r(\tau,N_I+1)=E_R(\tau,N_I)$.

\begin{acknowledgments} 
R.N. acknowledges partial support by the NSF under grant number
PHY-0854744.  
\end{acknowledgments}

\end{document}